# Multimodal Security of Iris and Fingerprint with Bloom Filters

**Dr. R. Sridevi[1] & P. Shobana[2]**

[1]*Associate Professor & Head, Department of Computer Science with Data Analytics, PSG College of Arts & Science, Coimbatore.*
[2]*Research Scholar, Department of Computer Science, PSG College of Arts & Science, Coimbatore.*
[1]*srinashok@gmail.com,* [2]*pshobana95@gmail.com*

***Abstract:*** *The standard methods of identification such as PIN Numbers (Personal Identification Number), Passwords, smart cards can be easily stolen and can be misused easily. To overcome this, biometric is introduced, as it will be unique to each individual. In this modern world, security has become a serious threat. So many biometric securities have also come up to secure. In biometrics, the iris has become more popular and widely used biometric because of its stability. This iris is researched for the past two decades along with computer technology. This paper will tell about basic concepts of iris, fingerprint, and securing iris with Cancelable Biometrics. Often we get iris templates and store in server or database. There is a chance of losing the data as well as security is a big question here. Here we are proposing a system for protecting iris templates using bloom filters with feature fusion of iris with fingerprint to provide security. Bloom filters are a useful asset in the fields of computer science. In feature-level fusion, the function units originating from more than one biometric assets are consolidated into a single feature set by the appliance of acceptable feature standardization, transformation, and reduction schemes. We can combine bloom filters with feature level fusion for more secure iris templates.*

**Keywords: Iris, Bloom filters, Feature level fusion, Biometric security**

## 1. Introduction:

Biometric is a more secure authentication method for systems where assets are given by finding an individual that is related to the specific measurements, dimensions, and characteristics of your body. For example, A person's iris, fingerprint, palm print, facial recognition. Unlike the conventional system, these characteristics cannot be stolen. Nowadays biometric is a common verification technique that is used over traditional verification. Generally, biometrics fall into two categories. 1) Physical 2) Behavioral. Physical identifiers are mostly unchanging and gadget-free. Physical identifiers include fingerprints, iris, palm print, facial recognition, retina. Behavioral identifiers are related to measurable patterns in human activities. This includes keystroke dynamics, gait analysis, signature analysis. Nowadays biometrics has made people's life easier. As we do not need to remember passwords, PINs.

In today's internet world, security is a serious threat. Biometrics are slowly changing conventional methods. Iris is one of the crucial biometric identifier, that is stable. Generally, these biometrics will be scanned and stored in a server or a database. If the server or database is hacked,  all the information will be stolen. So here we are using bloom filters and feature level fusion for the iris templates. The iris templates will be processed and features will be extracted.

The iris is a remotely noticeable, yet ensured organ whose one of a kind epigenetic design stays stable for the duration of grown-up life. These qualities make it exceptionally alluring for use as a biometric for distinguishing people. Picture handling methods can be utilized to remove the exceptional iris design from a digitized picture of the eye and encode it into a biometric layout, which can be put away in an archive database. The biometric format contains a target mathematical representation of one of a kind data in iris and it permits to do find similarities to be made between layouts. At the point when a subject wishes to be distinguished by iris acknowledgment framework, their eye is first shot, and then a format made for their iris region. This layout is then contrasted and different formats put away in a database until either a coordinating layout is found and the subject is recognized, or no match is found and the subject stays unidentified. John Daugman executed a working mechanized iris acknowledgment framework [18]. There are a few stages involved in iris recognition. They are Image capturing, Quality testing, Image preprocessing, Segmentation, Normalization, and Feature Extraction.

Image Capturing: The principal phase is to gather a huge database comprising of iris pictures from different people.

Quality testing: The subsequent stage is to check pictures for any noise.

Image preprocessing: If the image in the previous stage is not satisfied, then the iris image will be preprocessed to refine the quality of the iris.

Segmentation: This stage has to deal with the actual iris region from the digital image.

Normalization: After the segmentation process where the boundary of the iris is estimated, few techniques are used to analyze the texture.

Feature Extraction: This contains the texture of the iris. Algorithms are utilized to get the features.

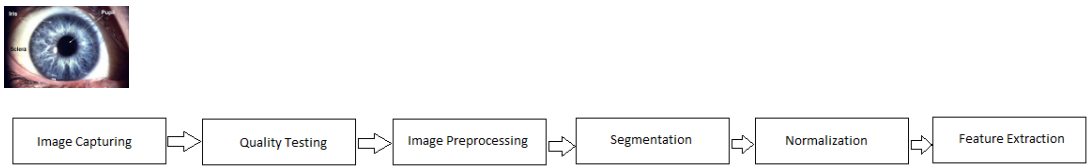

**Figure 1. Stages in Iris Recognition**

Fingerprinting is the most generally utilized and perceived biometric innovation for human confirmation.

Fingerprint sensing has become a common technique to be used in day to day life. A fingerprint is the portrayal of the dermal edges of a finger[17]. Dermal edges structure through a blend of hereditary and natural factors; the hereditary code in DNA gives general directions on the way skin should shape in a creating baby, however, the particular way it structures is an aftereffect of arbitrary occasions, for example, the specific situation of the embryo in the belly at a specific second. That's why even identical twins have different fingerprints[1]. Nowadays fingerprint recognitions are very accurate. There are a few stages involved in fingerprint recognition. They are Capturing, Enhancement, Feature Extraction.

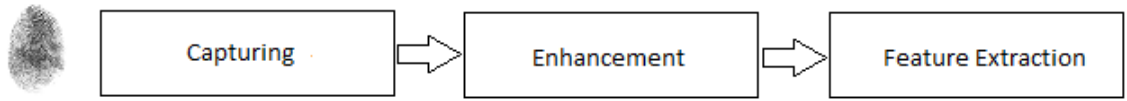

**Figure 2. Stages in Fingerprint Recognition**

## 2. Literature Survey:

Biometric format protection plans are normally classified as biometric cryptosystems and cancelable biometrics[2]. Cancelable biometric changes are structured such that it ought to be computationally difficult to recoup the original biometric information. The inborn quality (individuality) of biometric attributes ought not to be diminished applying changes while then again changes ought to be tolerant of intra-class variety. Also, the correlation of a few changed layouts must not uncover any data about the original biometrics (unlinkability).

A Bloom filter is a space-productive probabilistic information structure, formed by Burton Howard Bloom. To know Bloom channels, you initially need to comprehend hashing. To comprehend hashing. A hash resembles a unique finger impression for information. A hash work takes your information, which can be any length as information, and gives you back an identifier of a littler, fixed length, which you can use to record or analyze or recognize the information. The Bloom channel information structure tells whether a component might be in a set, or unquestionably isn't. The main potential blunders are false positives: a quest for a nonexistent component can offer a mistaken response. Bloom filters are both speedy and space saving. An unfilled Bloom channel is a bit array of m bits, all set to 0. There are likewise k distinctive hash works, every one of which maps a set component to one of the m bit positions.

To include a component, feed it to the hash capacities to get k bit positions, and set the pieces at these situations to 1. To know if an element is in the set, feed it to the hash capacities to get k bit positions. In the event that any of the pieces at these positions is 0, the component certainly isn't the set. On the off chance that all are 1, at that point the component might be in the set. Here is a Bloom channel with three added substances x, y, and z. It comprises of 18 bits and uses 3 hash capacities. The shaded bolts highlight the bits that the components of the set are mapped to.

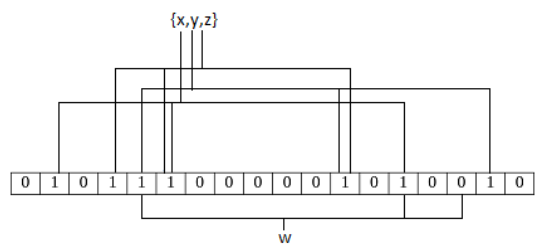

**Figure 3. Bloom Filter**

The component w is not there, because it hashes to a bit position having 0. Biometric indexing or filtering procedures are structured to diminish the quantity of competitor characters to be considered by an (iris) identification structure while scanning for a coordinate in an enormous archive of biometric reference information (templates)[3].

Unimodal biometrics is easy to build as it utilizes just a single quality or methodology of an individual yet it has numerous disadvantages like ridiculing, non-universality, etc. Multimodal biometrics utilizes more than one methodology of an individual like face, fingerprint, iris, retina, step, and so forth. Combining more than one attribute of an individual is troublesome thus these frameworks are safer and dependable. A biometric framework can be unimodal or multimodal. In the unimodal framework, anyone's attributes of an individual are taken and considered for validation. In multimodal frameworks, more than one characteristic of an individual is considered. The unimodal biometric frameworks had the impediment of noise into the information. The voice of a specific individual can be imitated, the marks can be effortlessly produced or fingerprints can be controlled, and so forth. In this way, the scientists concentrated on multimodal biometrics. In multimodal biometrics more than one attribute of an individual is taken so forging or controlling all the characteristics of a specific individual isn't simple thus the frameworks dependent on multimodal biometrics are safer and safe. In this, the information from various biometrics is caught and pre-handled. Then from the biometrics highlights are separated and coordinated with the stored formats in the database. There is a wide range of fusion procedures used for consolidating the various modalities used in the multimodal biometric system. A combination of these modalities is a significant and crucial step. Each of the combination levels has its favorable circumstances and drawbacks. The various modalities utilized in a multimodal biometric framework can be combined at various levels like element level, score level, and so forth. The levels of fusion are sensor level, feature level, score level, and decision level.

Sensor level: A few pictures are caught for the equivalent biometric characteristic and these pictures are intertwined to acquire a quality one[4].

Feature level: The highlights for every quality are separated and melded to make another dataset; the combination in this level is executed before performing coordinating calculation[4].

Score level: In this combination level, each biometric framework executes matching calculation to its extracted highlights independently then a combination is made between the matching outcomes [4].

Decision level: Each unimodal biometric framework settles on its choice independently then these individual choices are consolidated to produce the ultimate choice [4].

## 3. Proposed Work

### 3.1. Iris Recognition

The system consists of iris image obtaining, iris location and division, iris standardization, feature generation, and pattern identification. Here we use Fourier Descriptors for iris recognition[5]. Steps are given in Fig 4. Fourier Descriptors are used to create the component vectors by figuring coefficients of the changed iris picture in the recurrence area. The high-recurrence coefficients represent the data concerning the exact feature of the object while the low-recurrence coefficients represent the data concerning the overall qualities of the object. The Fourier Descriptors have been effectively utilized for some uses of the shape representation. The FDs have great attributes, such as it has resistance to noise, basic derivation, and straightforward standardization. FD is one of the best ones in perceiving the iris and then characterizing the human personality. With FD we can extract iris.

### 3.2. Fingerprint Recognition

The outcome and exactness of fingerprint acknowledgment rely upon the presence of valid minutiae. The algorithm is classified into two sections. Image Preprocessing and Post-handling. This algorithm upgrades the greater part of the phases of preprocessing for eliminating the noise, and make the clear cut fingerprint picture for feature extraction and improved the post-handling stages for dispensing with the false separated minutiae, to extract precise center point location, and coordinating valid minutiae[6]. This presents the post-preparing, as well as generally upgrade done in the fingerprint acknowledgment framework. The result of the preprocessing stage turns into a contribution for the post-preparing. The post-preparing includes particulars extraction, eliminate false details, center point discovery, and coordinating cycles. With this method, we can extract fingerprints.

### 3.3. Cancellable Biometrics

Generally, a cancellable biometric framework (CBS) is a component/signal-level format change approach, where the biometric property of a client is modified by boundaries got from either a client-specific secret key or key. Just the changed format is put away in the layout information base, and matching is performed inside the changing area.
Bloom filter has the advantage that space proficiency and question time effectiveness are far superior to standard calculations. A hash table can be utilized to decide if a component is in an assortment or not, and the recovery is exceptionally proficient.
Iris scanning and fingerprint scanning will be accomplished for the client to extract the features. After that, these two features will be merged into a single template. Then, this single template will be the input for bloom filters. Bloom filters will do some hashing and the template will be stored in the database.
First Iris and Fingerprint will be preprocessed and feature extraction will be done. Iris and fingerprint functions are fused. And hashing will be done on the fused template and can be stored in a database. Encryption: Here in encryption, iris and fingerprint will be taken from the user, and the features are extracted. After extraction, these two biological traits are fused to form a single one.
Biometric fusion(combination) is the term used to describe the system for incorporating information from at least two attributes. The fused feature will be given as input for the bloom filter. Bloom filter will do hashing for the given vectors and the output will be stored in the database.

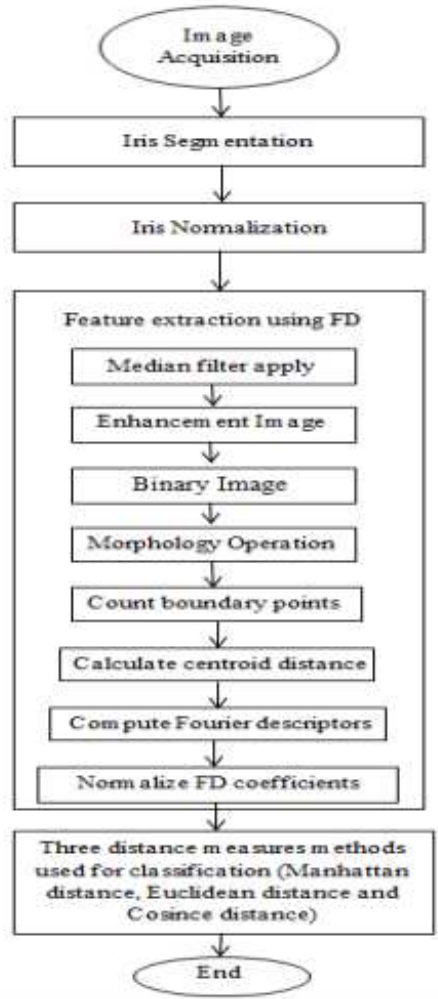

**Figure 4. Iris recognition system procedure using FD**

### 3.4. Decryption

Decryption is the oppposite of encryption. The user template will be matched with the template which is stored in the database. A stored template will be taken and decryption will take place on the template. The decrypted template will be compared with the user-provided template.

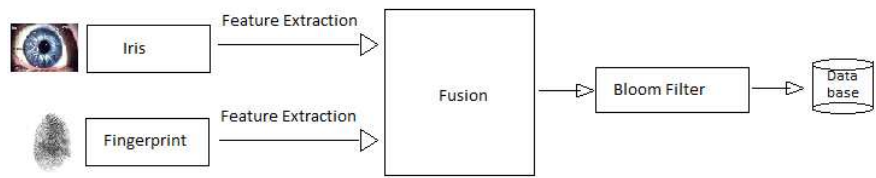

**Figure 5. Feature fusion and Bloom filter**

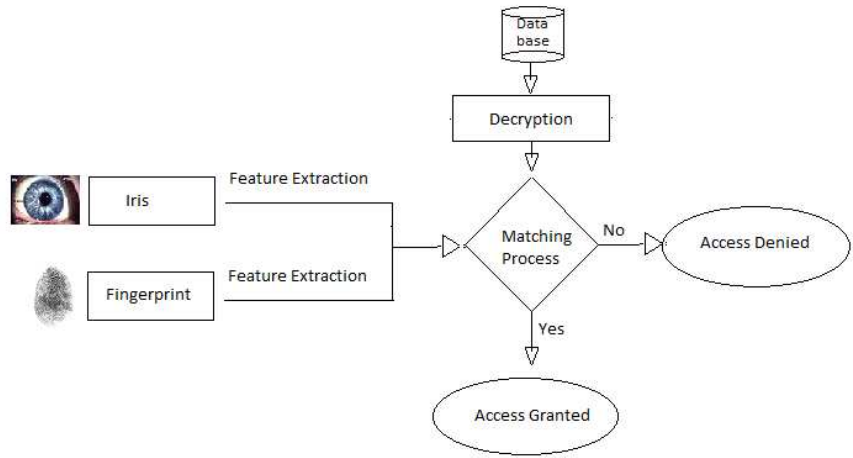

**Figure 6. Authentication**

## 4. Conclusion

This paper is all about securing biometric templates. Nowadays, Security is one of the biggest concerns in
all fields. Securing biometric sample is a serious threat. The protection of the individual physiological information should be shielded from being leaked. Numerous biometrics have come to secure. However biometric security is more secure than traditional systems. Cancellable biometrics are used to protect biometric data safely. Here we have used iris and fingerprint for feature level fusion. More than one biometric trait is used for feature fusion. Fusing more than one trait of an individual is difficult to forge and recreate. Numerous frameworks have been created which use multi modalities for security of the framework. Iris and fingerprint are extracted by using respective algorithms. Fingerprint and iris of a person will be fused utilizing feature-level fusion. Bloom filters are beneficial withinside the discipline of software. These bloom filters along with feature level fusion will give high security for the templates.

## REFERENCES

*5.1. Journal Article*

[1] Jain, A.K., Prabhakar, S., Pankanti, S.: On the similarity of identical twin fingerprints. Pattern Recognit. 35(11), 2653–2663 (2002).

[2] C. Rathgeb and A. Uhl, "A survey on biometric cryptosystems and cancelable biometrics," EURASIP Journal on Information Security, vol. 2011, no. 3, 2011

[3] R. Mukherjee and A. Ross. Indexing iris images. In ICPR, pages 1–4, 2008.

[4] Nidhi Srivastava, "Fusion Levels in Multimodal Biometric Systems– A Review ", International Journal of Innovative Research in Science, Engineering and Technology, Vol. 6, Issue 5, May 2017

[5]   Muthana H. Hamd and Samah K. Ahmed, "Biometric System Design for Iris Recognition Using Intelligent Algorithms",  International Journal of modern education and computer science, March 2018

[6]  Meghna B. Patel, Satyen M. Parikh and Ashok R. Patel, "An Improved Approach in Fingerprint Recognition Algorithm", Smart Computational Strategies: Theoretical and Practical Aspects pp 135-151, 2019

[7]  Dr.M.Gobi, Mrs.R.Sridevi,  "Performance Analysis of Biometric Image Encryption in Transformed Formats using Public Key Cryptography". International Journal of Scientific & Engineering Research(ISSN 2229-5518), Volume 6, Issue 2, February-2015.

[8]   Dr.M.Gobi, Mrs.R.Sridevi, "Multi-Biometric Authentication through Hybrid Cryptographic System", International Journal of Computing Algorithm (IJCOA)(ISSN:2278-2397), Volume: 04, Special Issue: March 2015, Pages: 1158 – 1161

[9] Mrs.R.Sridevi, S.Karthika, "Biometric Cryptosystem for VoIP Security using RSA Key Generation", International Journal of Software and Web Sciences (IJSWS), Issue 7, Volume 1(ISSN (Print): 2279-0063: ISSN (Online): 2279-0071) during February 2014

[10] Dr.M.Gobi, Mrs.R.Sridevi, " Biometric Security based on Reversible Data Hiding with RZL Code and Elliptic Curve Cryptography",  American International Journal of Research in Science, Technology, Engineering & Mathematics(ISSN (Online): 2328-3580), 10(2), March-May, 2015, pp.158-160

[11] Dr.R.Sridevi, Mrs.S.Selvi, "Progressing Biometric Security Concern with Blowfish Algorithm",  Scopus Journal - International Journal of Innovative Technology and Exploring Engineering (IJITEE), ISSN: 2278-3075, Volume-8, Issue- 9S2, July 2019. (Impact Factor: 5.54)

[12]  Dr.R.Sridevi, "Checking User Authentication by Biometric One Time Password Generation using Elliptic Curve Cryptography", International Journal of Computer Science & Engineering Technology (IJCSET), ISSN: 2229-3345, Vol. 8 No. 06, June 2017, pp. 213-218

[13] Dr.M.Gobi, Mrs.R.Sridevi, "An Implementation of Hybrid Cryptographic Protocol for Facial Image Security",  International Journal of current research in science and technology (ISSN: 2394-5745), Volume 1 Number 8 (2015).

[14] Dr.M.Gobi, Mrs.R.Sridevi, "A Design and Implementation of Hybrid Secured Protocol for Biometric Images", International Journal of Modern Trends in Engineering and Research (IJMTER) (ISSN (online): 2349-9745), Vol. 2, Issue 9, September 2015(impact factor: 1.711).

[15]  Dr.M.Gobi, Mrs.R.Sridevi, "A Secured Asymmetric Cryptosystem for Multi-Biometric Security", International Journal of Trend in Research and Development (IJTRD) (ISSN (online): 2394-9333), Vol. 2, Issue 5, September-October 2015(impact factor: 0.924).

[16] Mrs.R.Sridevi, S.Karthika "Secured Image Transfer Through DNA Cryptography Using Symmetric Cryptographic Algorithm", International Journal of Engineering Research and Science & Technology (ISSN 2319-5991) Vol. 4, No. 2, May 2015 Pages: 189 – 196, (impact factor: 3.785).

*5.2. Books*

[17] Maltoni, D., Maio, D., Jain, A.K., Prabhakar, S.: Handbook of fingerprint recognition. Springer, New    York (2003)

*5.3. Conference Proceedings*

[18] Daugman, J. 2002, "How Iris Recognition Works". Proceedings of 2002 International Conference on Image Processing".